\documentclass[12pt,a4paper,dvips]{article}

\newcommand{\be}{\begin{equation}}
\newcommand{\ee}{\end{equation}}
\newcommand{\bea}{\begin{eqnarray}}
\newcommand{\eea}{\end{eqnarray}}

\usepackage{graphicx}
\usepackage{amsbsy}
\usepackage{amssymb}

\begin{document}
\begin{center}
{\Large \bf  Wick Rotation and Abelian Bosonization\\}
\vspace*{1cm}
Aleksandar Bogojevi\' c\\
\emph{Institute of Physics, P.O.Box 57, Belgrade 11001, Yugoslavia\\}
\vspace*{5mm}
Olivera Mi\v skovi\' c\\
\emph{Institute of Nuclear Sciences ``Vin\v ca"\\
Department for Theoretical Physics, P.O.Box 522, Belgrade 11001, Yugoslavia\\}
\end{center}
\vspace*{5mm}
\begin{abstract}
We investigate the connection between Abelian bosonization in the Minkowski
and Euclidean formalisms. The relation is best seen in the complex time
formalism of S. A. Fulling and S. N. M. Ruijsenaars \cite{fulling}.
\end{abstract}
\vspace*{5mm}
\section{Introduction}

It is well-known \cite{mandelstam}-\cite{stone} that two-dimensional
free massless Dirac theory is equivalent to free massless scalar
field theory. In the operator formalism, this
equivalence is established by giving the following explicit construction
of fermionic fields in terms of bosonic fields:
\bea \label{explicit}
\psi_1 & = & \sqrt{\frac{c\mu}{2\pi}}\, \chi_1\, :e^{-i2\sqrt\pi \phi_1}:
\nonumber\\
\psi_2 & = & \sqrt{\frac{c\mu}{2\pi}}\, \chi_2\, :e^{i2\sqrt\pi \phi_2}:\, .
\eea
$\psi_{1/2}$ are two components of the spinor field $\psi$. On the
equation of motion $\psi_1=\psi_1(x^+)$, $\psi_2=\psi_2(x^-)$. In
Minkowski space $x^\pm$ are the light-cone coordinates, while in
Euclidean space we are dealing with holomorphic and anti-holomorphic
coordinates. Similary, on the equation of motion the scalar field $\phi(x)$
is a sum of a left-moving and right-moving part $\phi_1(x^+)$ and
$\phi_2(x^-)$. In (\ref{explicit}), $\mu$ is an infra-red cut off, 
necessary for dealing with massless scalar fields in two dimensions
\cite{klaiber}, while $c$ is a conveniently chosen constant
$\ln c=\Gamma'(0)$.

In the Minkowski formalism Mandelstam \cite{mandelstam}, \cite{coleman}
chose\footnote{ With this choice, the bosonic correlator
$$\langle \phi(x)\phi(y)\rangle = -\frac{1}{4\pi}\, \ln
c^2\mu^2 \left[ -(x-y)^2+i\varepsilon (x^0-y^0)\right] $$
automatically gives the correct fermi correlator and canonical
anti-commutators.} $\chi_1=\chi_2=1$. On the other hands, in the Euclidean
formalism \cite{stone}-\cite{neuberger},  it is
necessary to choose the $\chi$ prefactors to satisfy
\bea \label{prefactor}
\{ \chi_\alpha,\chi_{\beta\neq\alpha}\}&=&\{ \chi_\alpha,
\chi^\dagger_{\beta\neq\alpha}\}=0\nonumber\\
{[}\chi_\alpha,\chi_\alpha {]}&=&{[}\chi_\alpha,\chi^\dagger_\alpha {]}=0\, .
\eea
Without this choice, one indeed gets the correct fermi correlator,
but \emph{not}
the canonical anti-commutators. In most cases we are solely interested
in correlators (for example in conformal field theory) and the choice
of $\chi$ is not important. In this paper, however, we want to establish
the precise connection between the Minkowski and Euclidean formalisms. For
this reason it will be quite important to keep track of the $\chi$'s.


\section{Minkowski Formalism}

In his original paper, Mandelstam introduced a dual scalar field 
\be \label{intro}
\widetilde\phi(x)=\int_{-\infty}^{x^1}dy^1\, \dot\phi(x^0,y^1)\, ,
\ee
which satisfies $\partial_\mu\widetilde\phi={\varepsilon_\mu}^\nu
\partial_\nu\phi$. Both $\phi$ and
$\widetilde\phi$ are solutions of the massless Klein-Gordon equation $\partial^2
\phi=\partial^2\widetilde\phi=0$. The corresponding two-point
correlators are
\bea \label{corr-dual}
\langle\phi(x)\phi(y)\rangle &=&
\Delta_+(x-y)\nonumber\\
\langle\widetilde\phi(x)\widetilde\phi(y)\rangle &=&
\Delta_+(x-y)\nonumber\\
\langle\phi(x)\widetilde\phi(y)\rangle &=&
\widetilde\Delta_+(x-y)+\frac{i}{4}\nonumber\\
\langle\widetilde\phi(x)\phi(y)\rangle &=&
\widetilde\Delta_+(x-y)-\frac{i}{4}\, .
\eea
As in \cite{nakanishi}, we have introduced the auxiliary functions:
\bea \label{auxiliar}
\Delta_+(x) &\equiv& -\frac{1}{4\pi}\ln c^2\mu^2(-x^2+i\varepsilon x^0)\nonumber\\
\widetilde\Delta_+(x)&\equiv& -\frac{1}{4\pi}\ln \left|\frac{x^0+x^1-i\varepsilon}
{x^0-x^1-i\varepsilon}\right| \, .
\eea
The general solution of the equation of motion for the massless scalar field
is of the form $\phi(x)=\phi_1(x^+)+\phi_2(x^-)$, where we have introduced
light-cone coordinates $x^\pm=x^0\pm x^1$.
The left-moving and right-moving fields are determined up to a constant
(corresponding to the zero mode). Mandelstam chose the particular
decomposicion
\bea \label{decompos}
\phi_1 &=& \frac{1}{2}\, (\phi+\widetilde\phi)\nonumber\\
\phi_2 &=& \frac{1}{2}\, (\phi-\widetilde\phi)\, .
\eea
For later convenience let us list the scalar correlators in terms of the
above decomposition. We have
\bea \label{corr}
\langle\phi_1(x)\phi_1(y)\rangle &=&
\frac{1}{2}\, \left[ \Delta_+(x-y)+
\widetilde\Delta_+(x-y)\right] \nonumber\\
\langle\phi_2(x)\phi_2(y)\rangle &=&
\frac{1}{2}\, \left[ \Delta_+(x-y)-\widetilde\Delta_+(x-y)\right] \nonumber\\
\langle\phi_1(x)\phi_2(y)\rangle &=&-\frac{i}{8}\nonumber\\
\langle\phi_2(x)\phi_1(y)\rangle &=&\frac{i}{8}\, .
\eea
From (\ref{decompos}) we directly get the explicit bosonization formula
\bea \label{mandelstam}
\psi_1 &=& \sqrt\frac{c\mu}{2\pi}\, :\exp \left[ -i\sqrt\pi \phi(x)
-i\sqrt\pi\int_{-\infty}^{x^1}dy^1\, \dot\phi(x^0,y^1)\right] :\nonumber \\
\psi_2 &=& \sqrt\frac{c\mu}{2\pi}\, :\exp \left[ i\sqrt\pi \phi(x)
-i\sqrt\pi\int_{-\infty}^{x^1}dy^1\, \dot\phi(x^0,y^1)\right] :\, .
\eea
A trivial check shows that (\ref{mandelstam}) leads to the canonical
anti-commutation relations for $\psi$.

For later convenience we will write the above results in a different way.
Let us split the scalar field $\phi(x)$ into two pieces
\be \label{split}
\phi(x)=\varphi(x)+\rho(x)\, .
\ee
This split is chosen so that $\langle\varphi(x)\rho(y)\rangle=0$.
Our aim is to absorb all the $\varepsilon$-dependence of the two-point
correlators into the $\rho$ term. This is easily done, bearing in mind the
identity $\ln (a\pm i\varepsilon)=\ln |a| \pm i\pi \theta(-a)$, where $a$ is
a real number. As a result the $\rho$ correlators are
\bea \label{rho}
\langle\rho_1(x)\rho_1(y)\rangle &=&
\frac{i}{8}\, H(x^+-y^+)\nonumber \\
\langle\rho_2(x)\rho_2(y)\rangle &=&
\frac{i}{8}\, H(x^--y^-)\nonumber \\
\langle\rho_1(x)\rho_2(y)\rangle &=&-\frac{i}{8}\nonumber\\
\langle\rho_2(x)\rho_1(y)\rangle &=&\frac{i}{8}\, .
\eea
In these formulas
$H(x)$ is the Heaviside function $H(x)\equiv\theta(x)-\theta(-x)$.
Using (\ref{split}), our basic bosonization formula becomes
\bea \label{basic}
\psi_1 & = & \sqrt{\frac{c\mu}{2\pi}}\, :e^{-i2\sqrt\pi \phi_1}:~=~
\sqrt{\frac{c\mu}{2\pi}}\, \chi_1\, :e^{-i2\sqrt\pi \varphi_1}:\nonumber \\
\psi_2 & = & \sqrt{\frac{c\mu}{2\pi}}\, :e^{i2\sqrt\pi \phi_2}:~=~
\sqrt{\frac{c\mu}{2\pi}}\, \chi_2\, :e^{i2\sqrt\pi \varphi_2}:\, ,
\eea
where  the above procedure gives us the following explicit form
for the prefactors
\bea \label{chi}
\chi_1&\equiv& :e^{-i2\sqrt\pi \rho_1}:\nonumber \\
\chi_2&\equiv& :e^{i2\sqrt\pi \rho_2}:\, .
\eea
Let us emphasize that this is just Mandelstam's old result written in
different way --- in terms of the field $\varphi$ whose correlators do not
contain $\varepsilon$. For this reason, everything is consistent and we find
that the prefactors satisfy
\bea \label{algebra}
\{ \chi_\alpha(x),\chi_{\beta\neq\alpha}(y)\}&=&\{ \chi_\alpha(x),
\chi^\dagger_{\beta\neq\alpha}(y)\}=0\nonumber\\
{[}\chi_\alpha(x),\chi_\alpha(y) {]}&=&
{[}\chi_\alpha(x),\chi^\dagger_\alpha(y) {]}=0\, .
\eea
\section{Euclidean Formalism}

Bosonization in Euclidean space proceeds in much the same way as in
Minkowski space. The massless scalar field is
$\varphi(x)=\varphi_1(z)+\varphi_2(\bar z)$, where $z=x_0+ix_1$ and
$\bar z=x_0-ix_1$. These fields satisfy
\bea \label{corr-euc}
\langle\varphi_1(z)\varphi_1(z')\rangle&=&
-\frac{1}{4\pi}\ln c\mu (z-z')\nonumber\\
\langle\varphi_2(\bar z)\varphi_2(\bar z')\rangle&=&
-\frac{1}{4\pi}\ln c\mu (\bar z-\bar z')\nonumber\\
\langle\varphi_1(z)\varphi_2(\bar z')\rangle&=&0\, .
\eea
As we have stressed, in order to get the correct anti-commutation
relation for the fermi fields, one needs to introduce non-trivial
prefactors $\chi$ satisfying (\ref{prefactor}). From the previous
section, we see  that we have at our disposal an \emph{ explicit
construction} of these prefactors.
Turning that derivation on its head, we now introduce
$\rho(x)$ satisfying
\bea \label{corr-rho}
\langle\rho_1(z)\rho_1(z')\rangle&=&
-\frac{1}{4\pi}\ln \left[ 1-\frac{i\varepsilon}{c\mu (z-z')}\right] \nonumber\\
\langle\rho_2(\bar z)\rho_2(\bar z')\rangle&=&
-\frac{1}{4\pi}\ln \left[ 1-\frac{i\varepsilon}{c\mu (\bar z-\bar z')}
\right] \nonumber\\
\langle\rho_1(z)\rho_2(\bar z')\rangle&=&
-\frac{i}{8}\nonumber\\
\langle\rho_1(z)\rho_1(z')\rangle&=&
\frac{i}{8}\, .
\eea
This is just a straightforward extension to complex coordinates
of the Minkowski result given in (\ref{rho}). Finally, we make
a \emph{new} scalar field $\phi(x)=\varphi(x)+\rho(x)$.
The new field satisfies
\bea \label{corr-phi}
\langle\phi_1(z)\phi_1(z')\rangle&=&
-\frac{1}{4\pi}\ln c\mu (z-z'-i\varepsilon)\nonumber\\
\langle\phi_2(\bar z)\phi_2(\bar z')\rangle&=&
-\frac{1}{4\pi}\ln c\mu (\bar z-\bar z'-i\varepsilon)\nonumber\\
\langle\phi_1(z)\phi_2(\bar z')\rangle&=&-\frac{i}{8}\nonumber\\
\langle\phi_2(\bar z)\phi_1(z')\rangle&=&\frac{i}{8}\, .
\eea
We seem to have come to a strange result: the $i\varepsilon$-regularization
that is obviously necessory in Minkowski space saves the day in Euclidean
space as well.
\section{Complex Time Formalism}

Our intuition concerning the $i\varepsilon$ prescription is based on
the Feynman propagator. Based on that, $i\varepsilon$ wouldn't be
necessary in the Euclidean theory --- it doesn't do any damage, it is just
\emph{not necessary}. We shall, however, see that there exist objects 
for which
$i\varepsilon$ is needed in both Euclidean and Minkowski formalisms.
To do this, let us look at the following two-point functions
\bea \label{point}
\Delta_+(x-y)&\equiv&\langle\phi(x)\phi(y)\rangle\nonumber\\
\Delta_-(x-y)&\equiv&\langle\phi(y)\phi(x)\rangle=
{\Delta_+(x-y)}^*\nonumber\\
i\Delta_F(x-y)&\equiv&\langle T\phi(x)\phi(y)\rangle=
\theta(x^0-y^0)\Delta_+(x-y)+\theta(y^0-x^0)\Delta_-(x-y)\nonumber\\
i\Delta(x-y)&\equiv&{[}\phi(x), \phi(y){]}=\Delta_+(x-y)-\Delta_-(x-y)\, .
\eea
Let us next make an analytical continuation of these two-point functions
to complex time. Following Fulling and Ruijsenaars \cite{fulling},
\cite{mijic}, we take
\be \label{time}
t\to s=t+i\tau\, ,
\ee
and define the single function
\be \label{def}
{\cal D}(s)\equiv
\left\{ \begin{array}{ll}
\Delta_- (s)\, ,&\mathrm{Im}\, s>0\\
\Delta_+(s)\, ,&\mathrm{Im}\, s<0\, . \end{array}\right.  
\ee
This function is analytic in the $s$-plane except for a branch cut
along the $t$-axis\footnote{The fact that the cut is absent for
$|t|<|x|$ is a consequence of the causality condition which states that
 $[\phi ,\phi]$ vanishes outside the light-cone.} for $|t|>|x|$.
From this and (\ref{auxiliar}), it follows that we have
\be \label{de}
{\cal D} (s) = -\frac{1}{4\pi}\, \ln c^2 \mu^2 (-s^2+x^2)\, .
\ee
In terms of ${\cal D}(s)$, the standard Minkowski two-point functions
$\Delta_\pm$, $\Delta_F$ correspond to the contours $C_\pm$ and $C_F$
in Figure 1.
\begin{figure}[!ht]
    \centering
    \includegraphics[height=5.5cm]{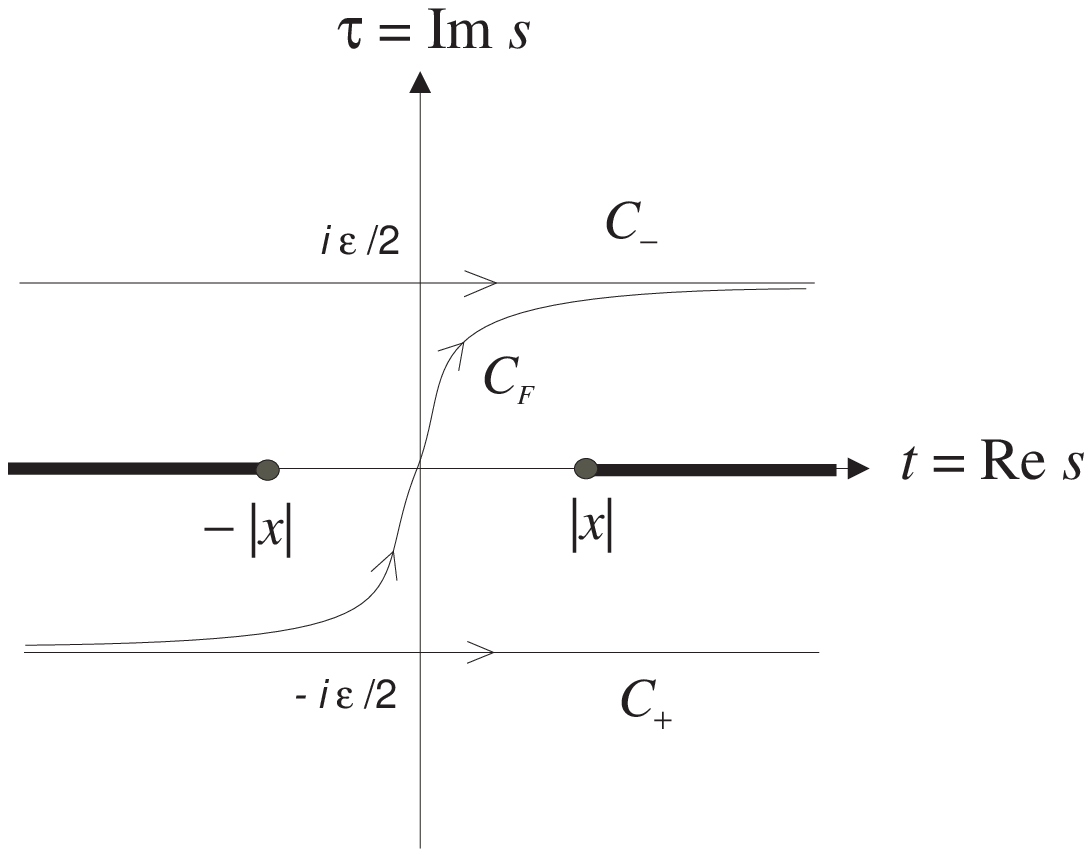}
    \caption{Contours corresponding to two-point functions $\Delta_\mp$
and $\Delta_F$.} 
\end{figure}
For example $\Delta_\pm(t)={\cal D}\left( t\mp\frac{i\varepsilon}{2}
\right)$.

Let us now look at the Euclidean theory.
The Euclidean propagator satisfies
\be \label{euclid}
\left( \partial^2_\tau+\partial^2_x\right)\Delta_E(\tau,x)=
-\delta(\tau)\delta(x)\, .
\ee
It can also be given in terms of ${\cal D}(s)$. This time the contour
is the imaginary $s$-axis. Wick rotation of the propagator corresponds 
to taking $C_F$ into $C_E$, as in Figure 2. 
\begin{figure}[!ht]
    \centering
    \includegraphics[height=5.5cm]{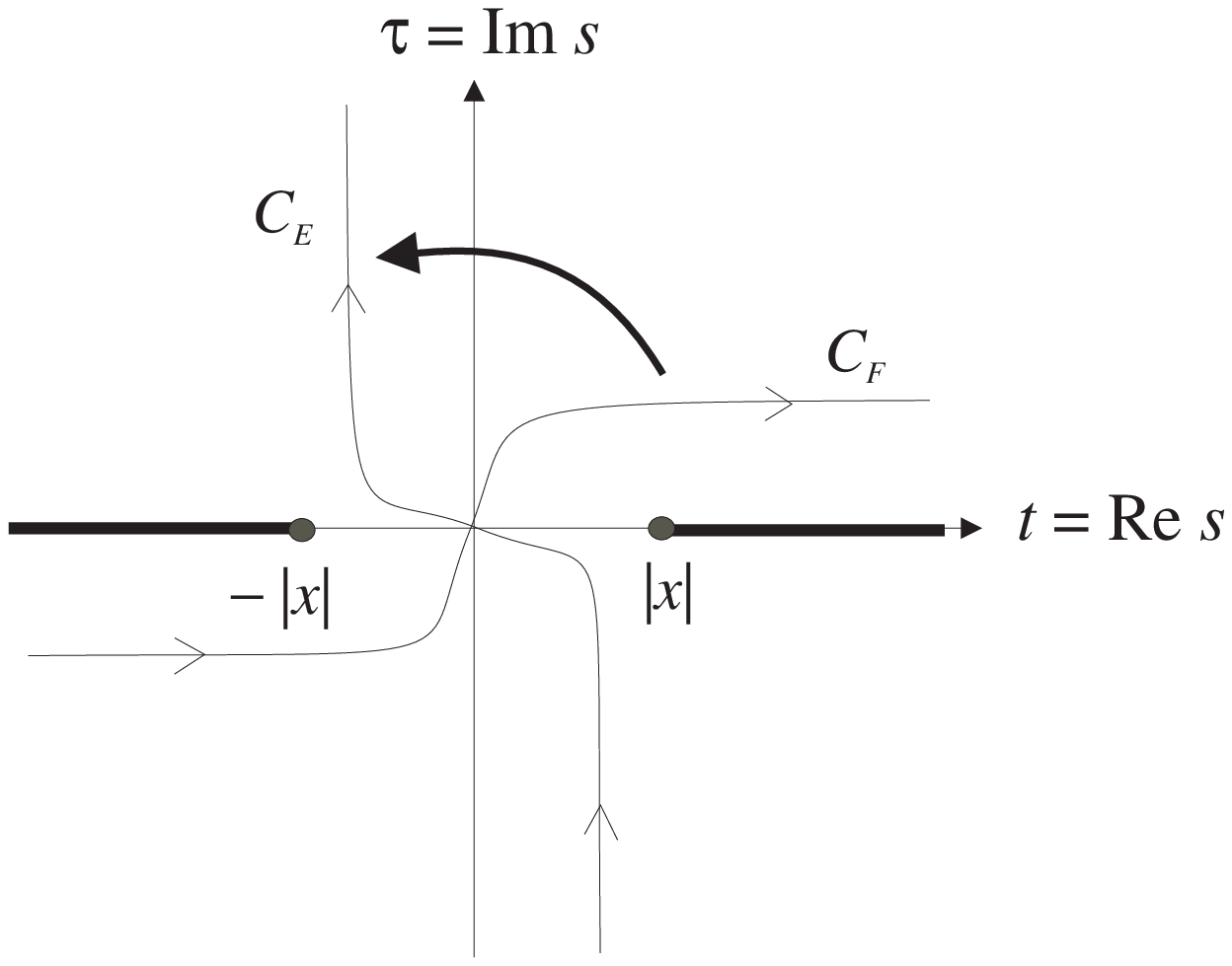}
    \caption{Wick rotation of the propagator: $C_F\to C_E$.} 
\end{figure}
Further, there are no 
obstructions, so $C_E$ can be deformed to the imaginary $s$-axis (hence 
there is no reason for $i\varepsilon$).

Let us next Wick rotate the $\Delta_-$ function. The contour $C_-$
rotates into $C_1\cup C_2$, as is shown in Figure 3. 
\begin{figure}[!ht]
    \centering
    \includegraphics[height=5.5cm]{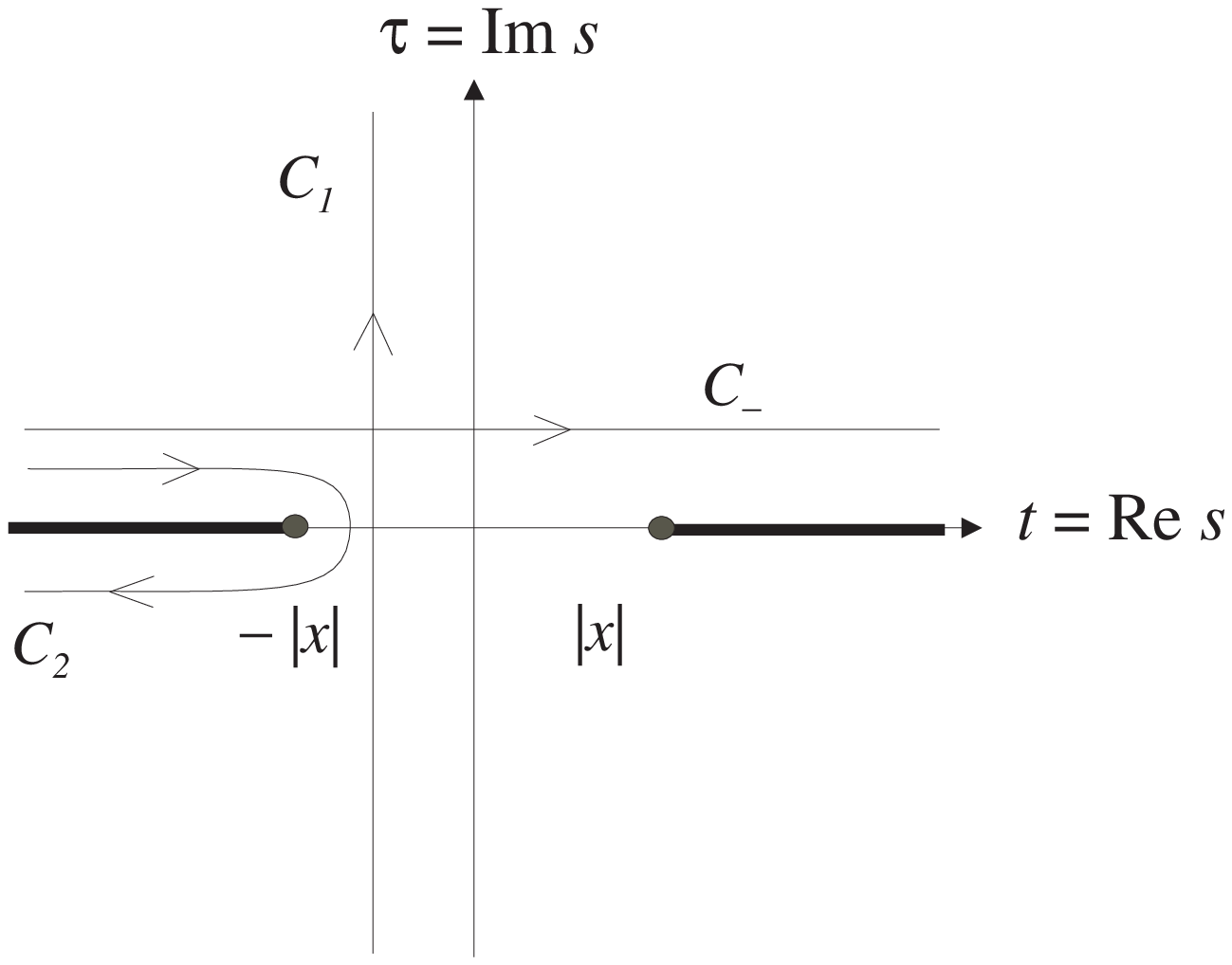}
    \caption{Wick rotation of $\Delta_-$ function: $C_-\to C_1\cup C_2$.} 
\end{figure}
As before, 
$C_1$ can be deformed to the
imaginary $s$-axis (it is not sensitive to $i\varepsilon$). However,
this time we have an additional piece corresponding to $C_2$.
From equations (\ref{point}) and (\ref{def}), it follows that this is
proportional to the canonical commutator. A naive Wick rotation of
$\Delta_-$ would not have seen this contribution. The $i\varepsilon$
prescription \emph{is} important in the whole complex time plane.

Let us emphasize this point once again. From the last relation in 
(\ref{point}), we see that the canonical commutator may be written
in terms of $\Delta_\pm$ as
\be \label{conclus}
{[}\phi(x), \phi(y){]}=\frac{1}{i}\, \left[ \Delta_+(x-y)-\Delta_-(x-y)
\right] \, .
\ee
Naive Wick rotation of the right hand side gives zero. On the other hand,
doing things carefully, \emph{i.e.} keeping track of the branch cut
(or equivalently of $i\varepsilon$), the right hand side of (\ref{conclus})
goes over into the correct canonical commutator of the Euclidean theory.
This is precisely the same thing that we saw in the previous section, where
we looked at bosonization in Euclidean space. The naive treatment gets wrong
results for fermi commutators --- they vanish. The correct fermi commutators
follow from incorporating the $i\varepsilon$ prescription.

\section{A Generalization}

At the end, let us look at a slight generalization of the standard
bosonization sheme along the lines of Nakanishi \cite{nakanishi}.
In his paper, Nakanishi looks at two scalar fields $\phi$ and
$\widetilde\phi$, connected by the duality relation
\be \label{duality}
\partial_\mu\widetilde\phi = {\varepsilon_\mu}^\nu\partial_\nu\phi\, .
\ee
Treating both fields as fundamental, we impose
\bea \label{fundament}
{[}\phi(x),\, \phi(y){]}&=&i\Delta(x-y)\nonumber\\
{[}\widetilde\phi(x),\, \widetilde\phi(y){]}&=&i\Delta(x-y)\, .
\eea
The mixed correlator is assumed to also give a $c$-number function.
From this and (\ref{fundament}), we find
\be \label{const}
[\phi(x),\widetilde\phi(y)]=i\widetilde\Delta(x-y)+i\alpha\, ,
\ee
where $i\widetilde\Delta\equiv\widetilde\Delta_+-
\widetilde\Delta_-$ and $\widetilde\Delta_-\equiv{\widetilde\Delta_+}^{^*}$.
The parameter $\alpha$ is a new undetermined real constant.
Mandelstam's bosonization sheme corresponds to the choice $\alpha = 1$.
Different authors have considered the cases $\alpha=\frac{1}{2}$
\cite{nakanishi}, and $\alpha=0$ \cite{neuberger},
\cite{suzuki}-\cite{hadjiivanov}.
Essentially, different choices for $\alpha$ correspond to different
boundary conditions. At the level of correlators, the general $\alpha$
case gives
\bea \label{general1}
\langle\phi(x)\phi(y)\rangle &=&
\Delta_+(x-y)\nonumber\\
\langle\widetilde\phi(x)\widetilde\phi(y)\rangle &=&
\Delta_+(x-y)\nonumber\\
\langle\phi(x)\widetilde\phi(y)\rangle &=&
\widetilde\Delta_+(x-y)+\frac{i}{4}\, \alpha\nonumber\\
\langle\widetilde\phi(x)\phi(y)\rangle &=&
\widetilde\Delta_+(x-y)-\frac{i}{4}\, \alpha\, ,
\eea
or, equivalently, using (\ref{decompos}) we find
\bea \label{general2}
\langle\phi_1(x)\phi_1(y)\rangle &=&
\frac{1}{2}\, \left[ \Delta_+(x-y)+
\widetilde\Delta_+(x-y)\right]\nonumber\\
\langle\phi_2(x)\phi_2(y)\rangle &=&
\frac{1}{2}\, \left[ \Delta_+(x-y)-\widetilde\Delta_+(x-y)\right]\nonumber\\
\langle\phi_1(x)\phi_2(y)\rangle &=&-\frac{i}{8}\, \alpha\nonumber\\
\langle\phi_2(x)\phi_1(y)\rangle &=&\frac{i}{8}\,\alpha \, .
\eea
The above authors do not look at fermi commutators. A simple calculation, as
in the first section, shows that one gets the correct results for
$\{ \psi_1,\psi_1\}$ and $\{ \psi_2,\psi_2\}$ for all values of $\alpha$.
On the other hand, the mixed commutator $\{ \psi_1,\psi_2\}$  is found
to be proportional to $1+{(-1)}^\alpha$. Therefore, by insisting on canonical
fermi commutators, we impose the condition $\alpha=2n+1$, where $n$ is an
integer. Note that Mandelstam's bosonization scheme satisfies
this condition, while the others do not.

The issue of boundary conditions in bosonization is a very important one.
It is still an open problem. We are currently looking into that on the
example of the bosonization of the massive Thirring model.

\newpage

\end{document}